\documentclass[12pt,preprint]{aastex}

\newcommand{\Bamb}{B_{amb}^{}}
\newcommand{\rhotop}{\rho_{top}^{}}

\newcommand{\degree}{\ensuremath{^\circ}}

\slugcomment{Submitted to ApJ Letters, 9th Nov 2007}

\shorttitle{Jets in coronal holes: observations and modelling}
\shortauthors{Moreno-Insertis et al.}

\begin{document}

\title{Jets in coronal holes: Hinode observations and 
  3D computer modelling}

\author{F. Moreno-Insertis\altaffilmark{1},  K. Galsgaard\altaffilmark{2},
  I. Ugarte-Urra\altaffilmark{3}} 

\altaffiltext{1}{Instituto de Astrofisica de Canarias (IAC), 38200 La Laguna
  (Tenerife), Spain; fmi@iac.es}

\altaffiltext{2}{Niels Bohr Institute, University of Copenhagen, Copenhagen,
  Denmark; kg@astro.ku.dk}

\altaffiltext{3}{Space Science Division, Naval
  Research Laboratory,   Washington DC 20375, USA; iugarte@ssd5.nrl.navy.mil}

\begin{abstract}
Recent observations of coronal hole areas with the XRT and EIS instruments
onboard the {\it Hinode} satellite have shown with unprecedented detail the
launching of fast, hot jets away from the solar surface. In some cases these
events coincide with episodes of flux emergence from beneath the
photosphere. In this letter we show results of a 3D numerical experiment of
flux emergence from the solar interior into a coronal hole and compare them
with simultaneous XRT and EIS observations of a jet-launching event that
accompanied the appearance of a bipolar region in MDI magnetograms.  The
magnetic skeleton and topology that result in the experiment bear a strong
resemblance to linear force-fee extrapolations of the {\it SOHO}/MDI
magnetograms. A thin current sheet is formed at the boundary of the emerging
plasma. A jet is launched upward along the open reconnected field lines with
values of temperature, density and velocity in agreement with the XRT and EIS
observations. Below the jet, a split-vault structure results with two
chambers: a shrinking one containing the emerged field loops and a growing
one with loops produced by the reconnection.  The ongoing reconnection leads
to a horizontal drift of the vault-and-jet structure. The timescales,
velocities, and other plasma properties in the experiment are consistent with
recent statistical studies of this type of events made with {\it Hinode} data.
\end{abstract}


\section{Introduction}\label{sec:introduction}

\begin{figure}[htbp!]
\centering
 \includegraphics[width=8cm]{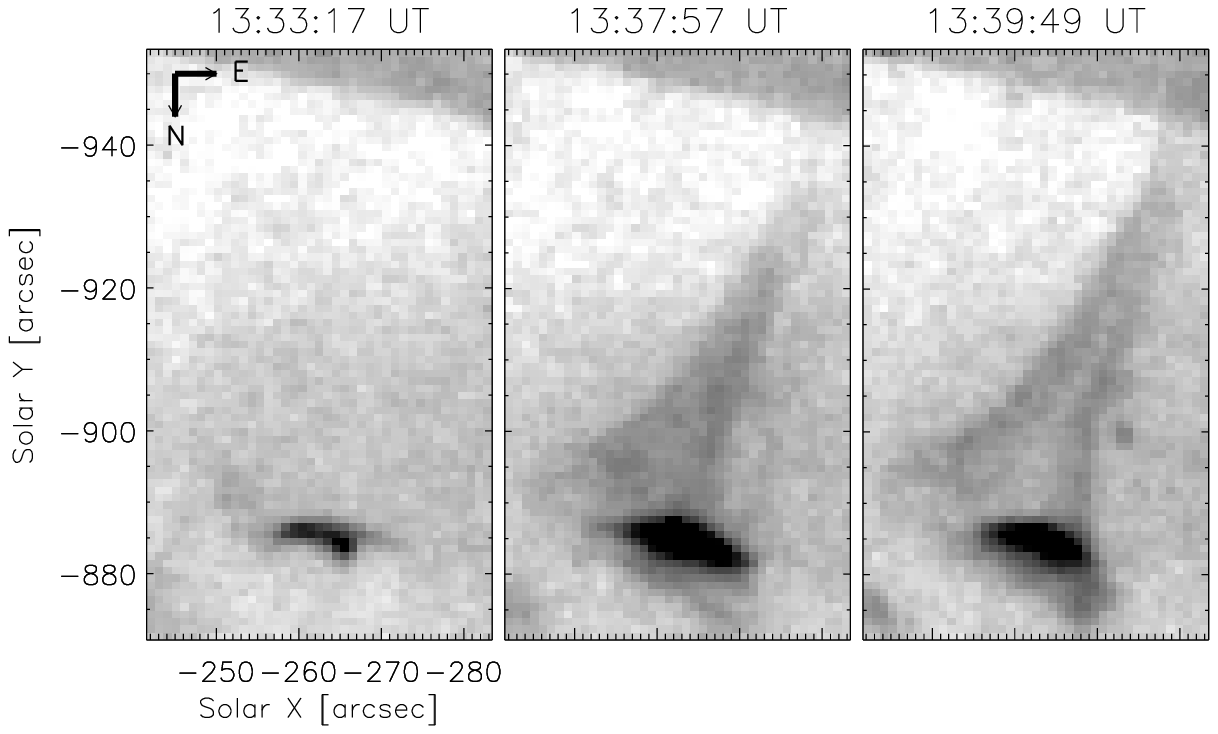}
\caption{Polar coronal jet observed with XRT Al\_poly/Open filter combination
on the South coronal hole on January 20 2007.  Color scale is reversed.}
\label{fig:polarjet}
\end{figure}

X-ray jets in coronal holes are being observed with unprecedented quality by
the instruments onboard the {\it Hinode} satellite, in particular the X-Ray
Telescope (XRT) \citep{golub2007} and the Extreme UV Imaging Spectrometer
(EIS) \citep{culhane2007}.  With their enhanced sensitivity and spatial
resolution, those instruments have revealed that the frequency of jet
formation is at least an order of magnitude greater than was thought
previously \citep{cirtain2007}. The new observations are leading to an
in-depth revision of the statistics of jet properties \citep{Shimojo+ea96,
savcheva2007}.  Sometimes, the formation of the jet is seen to occur
simultaneously with an episode of flux emergence from the solar interior
(e.g., \citealt{canfield1996, jiang2007}).  In many XRT events, a brightening
close to the surface is followed by enhanced emission in a region with an
{\it inverted-Y} shape above the former (Fig.~\ref{fig:polarjet}). The {\it
  tail} of the inverted {\it Y} coincides with a fast jet.

This kind of jets is most probably a consequence of the field line reconnection
between the upcoming magnetic system and the open
ambient field in the coronal hole, as originally suggested by theory and 2D
numerical experiments \citep{heyvaertsetal77, yokoyama_shibata_96}.  In this
letter, we analyze XRT and EIS observations of a jet in a coronal hole
cospatial with a simultaneous flux emergence event observed with {\it
  SOHO}-MDI (Sec.~\ref{sec:observations}).  We study the magnetic topology
using linear force-free extrapolations from the MDI data. We then present 
the results of a three-dimensional experiment of flux emergence into a region
in the atmosphere with temperature, density and magnetic field values akin to
those in a coronal hole (Sec.~\ref{sec:num_exp}). We explain the resulting three-dimensional
structure (geometry, topology, dynamics) and show the excellent overall
agreement with the observations.

\section{Observations}\label{sec:observations}
\begin{figure}[htbp!]
\centering
\includegraphics[angle=0,width=10cm]{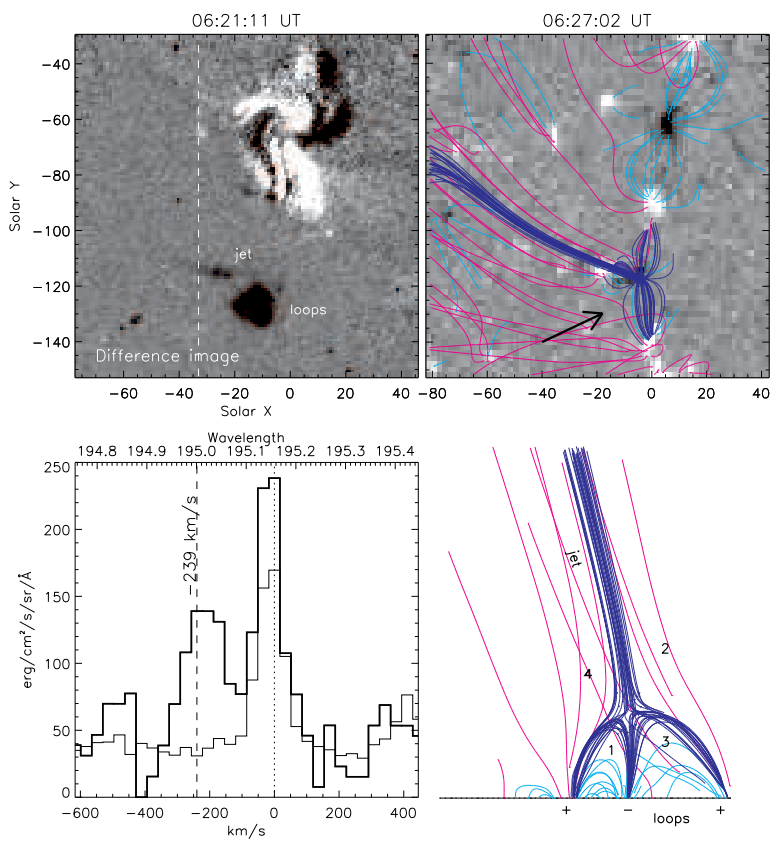}
\vskip 4mm
\caption{Coronal jet observed in an equatorial coronal hole on March 10 2007.
Top left panel: difference XRT Open/Ti\_poly image of the jet; the dashed
line indicates the EIS slit position.  Top right: MDI magnetogram and linear
force free magnetic field extrapolation ($\alpha=-0.007$ Mm$^{-1}$); the
color code is: a) dark blue: field lines near the apparent magnetic null
and the separatrices; b) light blue: field lines with two foot-points in the
photosphere; c) red lines: open field lines. Bottom right: side view as
indicated by the black arrow in the panel above. Numbers indicate
connectivity regions. Bottom left: 
\ion{Fe}{12} 195.1 \AA\ spectral profile at the jet location (thick line) at
06:21 
UT. The thin line corresponds to an average profile elsewhere in the coronal
hole. }
\label{fig:equatorialjet}
\end{figure}

A large coronal hole was observed near disk center on March 10, 2007 by the
{\it Hinode} satellite \citep{kosugi2007}. We present EIS and XRT observations of a
jet that appeared at 06:20 UTC at solar coordinates ($-25\arcsec,
-115\arcsec$) within the equatorial hole, at the same time as a magnetic
bipole emerged at the photosphere (Fig.~\ref{fig:equatorialjet}). A
low-latitude event was chosen to facilitate the extrapolation and radial
velocity measurement.  The evolution of the jet was inspected using XRT
$512\times512$ pixel images taken in the Open/Ti\_poly filter combination at
a cadence of 70 s. Spatial sampling is 1$\arcsec$ per pixel.  The EIS
observations consisted of 30 s exposures in sit-and-stare mode with the
$1\arcsec\times512\arcsec$ slit.  Standard routines were used in the data
processing.  The alignment between both instruments was done via
cross-correlation of North-South cross-sections of the X-ray images with the
\ion{Fe}{12} emission along the EIS slit.  Accuracy is down to between
$1\arcsec$ and $2\arcsec$ in the N-S direction and less than 5$\arcsec$ in
the E-W.  For the analysis of the magnetic flux density we used full disk
magnetograms from {\it SOHO}-MDI \citep[Michelson Doppler
  Imager,][]{scherrer1995}. Only magnetograms with a 96-min cadence were
available at the time of the jet. The closest one in time (06:27 UTC) shows a
newly emerged bipole under the location where the jet is formed. The magnetic
flux density of the emerged polarities is in the range (-70, 88) Mx cm$^{-2}$
[i.e., $(-7,\, 8.8)\,10^{-7}$ Wb m$^{-2}$] with total fluxes of $\approx
1\times10^{19}$ Mx $= 1\times10^{11}$ Wb.  The magnetic
topology was determined with a linear force-free extrapolation method
\citep[see details in][]{ugarte-urra2007} and reveals the existence of a
coronal null point and a separatrix dome that encloses the connectivity
domain of the newly emerged flux.

In the XRT images (as, e.g., in the top-left panel of
Fig.~\ref{fig:equatorialjet}), we first see the appearance of small coronal
loops that connect the pre-existing dominant ambient positive flux to the
newly emerged negative polarity. This is followed immediately by the
formation of the jet, oppositely oriented and rooted in the positive flux,
with the footpoints most likely located where the separatrix surface
intersects the photosphere. The top-right panel of the figure shows the
relationship of the loops and jet to the magnetic topology from the
extrapolation.  The bottom right panel shows a side view of the topology with
four labeled connectivity domains.  In region 1 lie the magnetic connections
between the two newly emerged polarities. The loops on the XRT image
correspond to region 3. A natural explanation for this configuration is that
the emerging flux (region 1) is merging with flux from region 2 yielding two
reconnected-line regions (3 and 4). The jet must be in an open-field region,
so it must be in region 4, and close to the separatrix (dark-blue field
lines). This is consistent with what the projected extrapolation and the XRT
image suggest and is also in agreement with the results of the numerical
simulation presented in Sec.~\ref{sec:num_exp}.  At least two consecutive jet
events can be identified in the XRT movie before the loops grow and finally
populate the whole enclosed domain.  The initial and most impulsive phase
lasts around $12$ min.

The EIS slit crosses the jet structure at about 15$\arcsec$ to the East
side of its origin.  EIS's unique combination of high spatial, spectral
and temporal resolution can provide diagnostics for the density and
velocity (for an in-depth discussion of previous EUV spectroscopic
observations of jets see \citealt{wilhelm2002, ko2005,
pike_harrison_1997} and references therein).  On the jet we measured
(Fig.~\ref{fig:equatorialjet}, bottom-left panel) a line-of-sight
blue-shifted velocity of up to $240$ km s$^{-1}$ in the \ion{Fe}{12}
195.12-\AA\ line. The zero value is given by an average profile in the
quiet coronal hole.  After averaging two pixels along the slit and two
exposures, to increase the signal-to-noise ratio, and subtracting the
background emission, we also obtained an estimate for the electron
density through the line ratio \ion{Fe}{12} 195.1/186.9:
$5.5\times10^9\rm{cm}^{-3}$. The density estimate is consistent with a
filling factor of $\approx\hskip -2pt 0.2$ for the \ion{Fe}{12}
195.1\AA\ column emission measure, assuming photospheric abundance
\citep{grevesse2007}, a temperature corresponding to the maximum in the
ionization fractions \citep{mazzotta1998} and a column depth of $4.7$
Mm, equal to the jet's measured width.  We used CHIANTI v5.2
\citep{landi2006} for the calculations.

\section{Numerical experiment}\label{sec:num_exp}

The numerical results were obtained for a three-dimensional domain of size
$34 \hbox{\ Mm} \times 38 \hbox{\ Mm}$ in the horizontal directions and $33$
Mm in the vertical direction, $z$, $29$ Mm of which are above the
photosphere. For the unmagnetized hydrostatic background at time $t=0$ we
chose an adiabatic stratification in the solar interior and a domain at
photospheric temperature in the first $2$ Mm above the surface; above it, we
have a steep temperature gradient mimicking the transition region and,
finally, an isothermal ($T=1.1\times10^6$ K) corona in the topmost $25$ Mm
(as in Fig.~1 of \citealt{archontis_etal_04}, but with a much larger vertical
domain).  In the main experiment described here, $\rhotop$, the initial
coronal density at the top of the box, was $5 \times10^{-16}$ g cm$^{-3}$, 
equivalent to an atom number density of $\approx 2\times10^{8}$ cm$^{-3}$. The
domain is endowed at time $t=0$ with a uniform magnetic field contained in
the $xz$ plane of strength $\Bamb = 10$ G $= 10$ mT and pointing $25\degree$
away from the vertical.  This field is dynamically dominant in the corona
[$\beta \approx \hbox{O}(10^{-2})$], whereas it is very weak ($\beta \gg 1$)
in the solar interior. Density and field strength are adequate to simulate
coronal holes (see \citealt{wilhelm06, harvey_recely_2002}).  To explore the
parameter space, experiments have also been carried out for $\Bamb$ equal to
$25$ G and $5$ G, for $\rhotop = 1.5\times 10^{-14}$ g cm$^{-3}$, and for
inclinations of $15\degree$ and $0\degree$ to the vertical. We solve the MHD
equations and assume an
ideal gas law, no radiation cooling, and no heat conduction. The
system is solved using the staggered-grid MHD code of our previous
experiments (\citealt{nordlund+galsgaard97}; see
\citealt{archontis_etal_04}). The numerical grid is non-uniform in the
vertical direction, with minimum resolution of $5.3$ points per scale height
(pps) in the photosphere ($65$ pps in the corona), and has $350 \times 322
\times 320$ nodes in the $(x,y,z)$ directions.

The emergence process is initiated by including a twisted magnetic flux tube
with axis along the $y$ coordinate direction located $1.7$ Mm below the
photosphere and with maximum field strength $3.8$ kG.  The tube is endowed
with buoyancy in its central part so that it rises and reaches the surface in
about $15$ min. The main features of these initial phases are as
described in our previous experiments (e.g., \citealt{archontis_etal_04,
archontis_etal_05}).  Once in the atmosphere, the rising plasma has a
considerable magnetic pressure excess and expands in all directions, albeit
preferentially horizontally, whereby it adopts a {\it helmet} shape
protruding from the solar interior. The pressure of the ambient coronal field
opposes the expansion and the emerged volume reaches
a maximum size of about $8$ Mm in height, $11$ Mm in the $x$
direction (i.e., transverse to the original tube axis) and $17$ Mm in the $y$
direction.

\begin{figure}[htbp!]
\centering
\includegraphics[width=8.6cm,height=18.75cm]{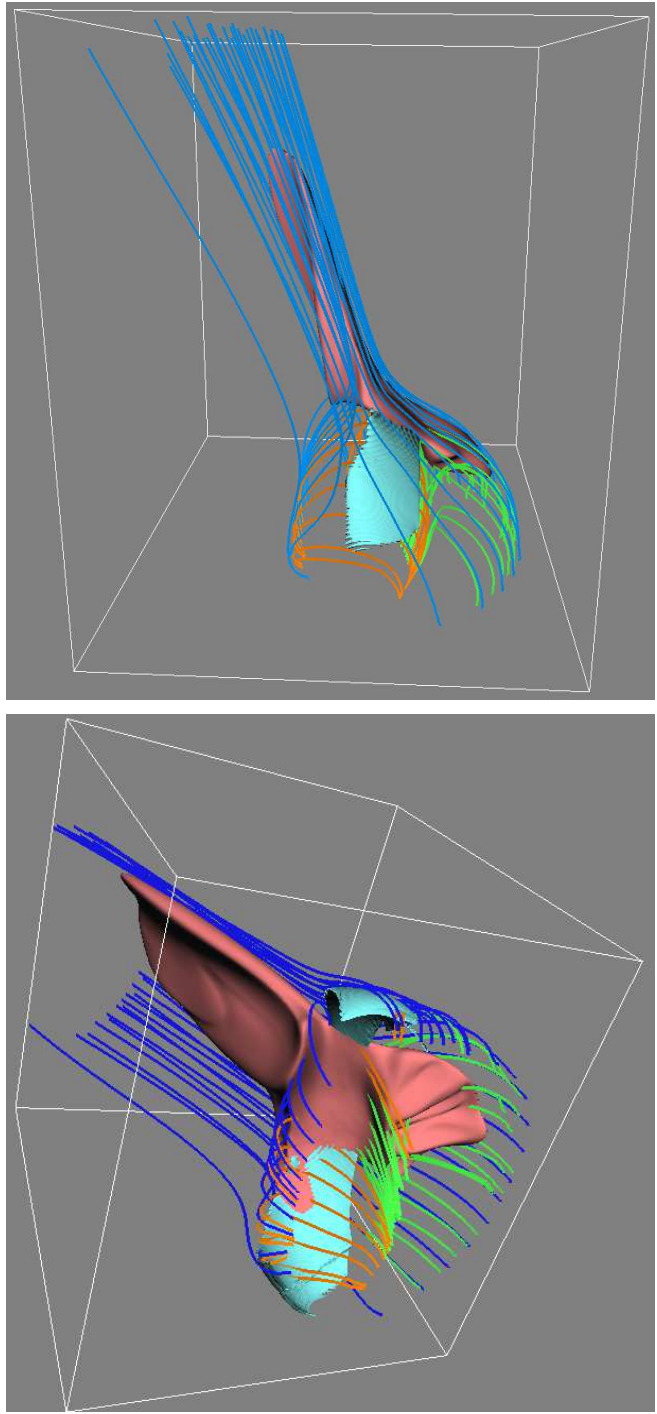}
\caption{Three-dimensional view of the emerged region, the reconnection site
    and the jet at the time of peak reconnection activity (top: side
    view; bottom: view from below). The isosurface of $j/B$ (in blue)
    delineates the collapsed current sheet on the side of the emerged volume.
    The temperature isosurface ($T=6.5\times10^{6}$ K; in red) encompasses
    both the reconnection site and the jet volume. Underlying the jet and
    current sheet, a double set of current loops (emerged and reconnected,
    with field lines in orange and green, respectively) is visible, giving a
    'double-chambered' vault shape to the region below the jet.  }
\label{fig:3D_plots}
\end{figure}

\begin{figure}[htbp!]
\centering
\includegraphics[width=8.7cm,height=14.6cm]{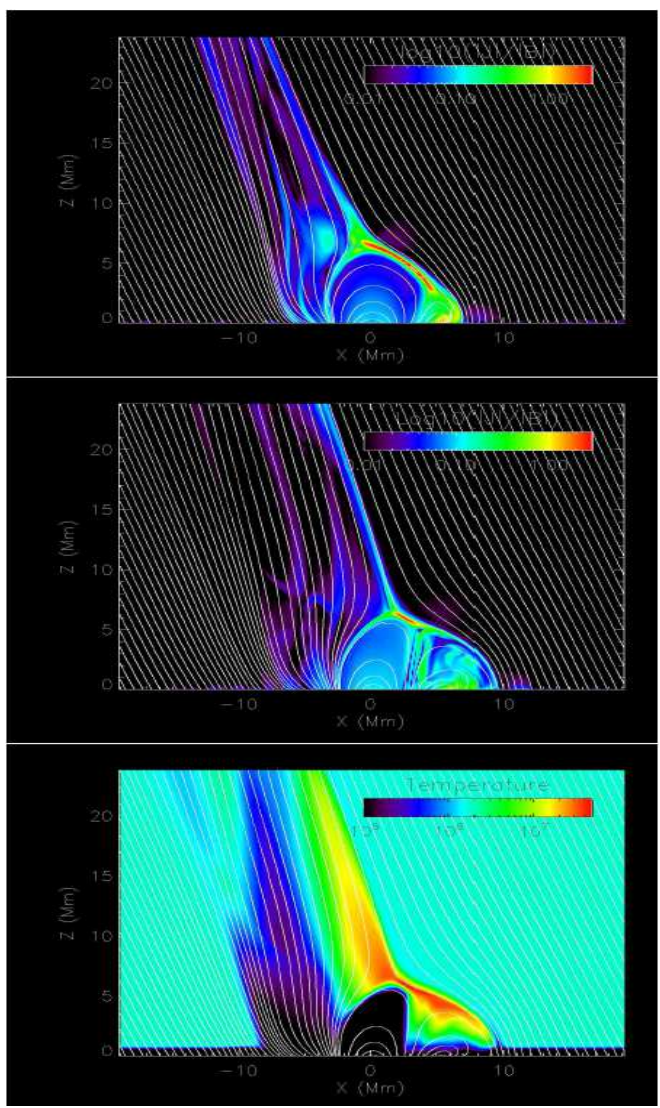}
\caption{Vertical cuts at different stages of the evolution. Top: $j / B $
  distribution and field line projection at an early stage. A thin current
  sheet is situated to the top-right of the emerged coronal loops. The
  diffuse, elongated current perturbations to the top and left of the emerged
  volume correspond to previously reconnected field lines. Center: total
  velocity map and field line projection at the peak activity phase of the
  jet, occuring about $7$ min later than the top panel. The double-chambered
  structure below the jet is clearly visible. Bottom: temperature
  distribution for the same snapshot as the central panel.  An inverted-Y
  structure appears prominently, with $T\hskip -3pt \approx \hskip -3pt 3\,10^7$
   K (reconnection site) and $\approx\hskip -3pt 10^7$ K (upward
  pointing jet). }
\label{fig:2D_plots}
\end{figure}

The expansion just described firmly presses against each other the magnetic
field in the emerged volume and the ambient corona. The field in the rising
plasma is twisted around the tube axis; hence, it is almost counteraligned
with the coronal field on one of the sides of the emerging volume.  A thin,
elongated current sheet is thereby formed (blue isosurface in Fig.~\ref{fig:3D_plots})
that embraces that side of the volume, resembling those obtained by
\citet{archontis_etal_05}.  Seen in a vertical cut, the sheet appears as a
thin stripe (in red in Fig.~\ref{fig:2D_plots}, top panel) of Syrovatskii
type. Reconnection takes place across the current sheet: the arch-like loops
next to it inside the {\it helmet} merge with the open loops coming in from
the other side of the sheet. Two regions of reconnected field result;
at the thin edge at the top of the current sheet, a set of open field lines
are ejected: that region is visible in Fig.~\ref{fig:2D_plots} (top panel) as
a vertical band of diffuse current perturbations.  At the lower edge, closed
coronal loops ensue. 

High velocity outflows are ejected from the upper and lower edge of the
current sheet.  The outflow at the upper edge (Fig.~\ref{fig:2D_plots},
center) is roughly horizontal and reaches peak speeds around $400$ km
s$^{-1}$ (of order the local Alfv\'en speed). Shortly after leaving the
reconnection site, it is deflected into two secondary jets propagating upward
and downward along the field lines, basically as described in 2D by
\citet{yokoyama_shibata_96}.  The upgoing branch attains vertical velocities
above $200$ km s$^{-1}$ (Fig.~\ref{fig:2D_plots}, center).  High temperature
values are reached in the jets and in the reconnection site itself. At the
peak of activity (Figs.~\ref{fig:3D_plots} and \ref{fig:2D_plots}), the
$6.5\times10^{6}$ K isosurface (in pink in Fig.~\ref{fig:3D_plots}) extends
over the current sheet and into the jet.  Values as high as $3\times10^7$ K
(reconnection site) and slightly above $10^7$ K (jet) are reached
(Fig.~\ref{fig:2D_plots}, bottom).  The high-T regions have the shape
of an inverted-Y on top of the emerged material (Figs.~\ref{fig:3D_plots} and
\ref{fig:2D_plots}), not unlike those observed by {\it Hinode}
(Fig.~\ref{fig:polarjet}).  The density of the jet at the peak of activity is
about $10$ times the density of the unperturbed corona.

As reconnection proceeds, a {\it double-chambered vault} structure of closed
loops develops below the jet, visible below the blue stripe in
Figs.~\ref{fig:3D_plots} and \ref{fig:2D_plots} (center and bottom panels):
in one of the chambers ({\it the emerged chamber}), coronal loops with the
original connectivity are still present; in the other ({\it the reconnected
chamber}), the new set of closed, high-T coronal loops are being stored. The
whole system strongly resembles the topology obtained in
Sect.~\ref{sec:observations} through extrapolation from MDI data. As time
proceeds, the {\it emerged chamber} decreases in transverse size while the
{\it reconnected chamber} grows to a size 
similar to the original emerged volume.  There results an
apparent sideways {\it drift} of the vault and of the jet in the direction
toward the reconnected loops, with roughly $10$ km s$^{-1}$ drift speed.
This may correspond to the drift discussed by \citet{savcheva2007}.

The main phase of reconnection with jet speeds above $100$ km s$^{-1}$
lasts around $7$ min, followed by an extended phase with lower temperature
and jet velocities, for a total duration of some $20$ min. In the early
stages of reconnection, in turn, we observe  cool and dense plasma
being ejected from the reconnection site. At that stage, the high-density
shell that covers the emerging plasma is being merged with the
ambient coronal field.  The resulting reconnected open field lines become
loaded with high-density [$\rho \approx \hbox{O}(10^2 \rhotop)$], low-T
[O($10^5$) K] material (visible to the left of the hot jet in
Fig.~\ref{fig:2D_plots}, bottom panel).

\section{Discussion}\label{sec:discussion}

In this letter we have shown that the inverted-Y shaped jets recently
observed by {\it Hinode}/XRT in coronal holes are a natural consequence of the
emergence of magnetic flux from below the surface and its interaction with
the preexisting open field in the coronal hole. Flux emergence is a prime
candidate to trigger reconnection and the consequent launching of jets along
field lines in the corona, as proposed through theoretical or 2D numerical
results \citep{heyvaertsetal77, yokoyama_shibata_96} and through 3D
experiments of jet formation in a horizontally magnetized atmosphere
\citep{archontis_etal_05, galsgaardetal07}. Our present 3D experiments had a
field configuration and stratification parameters akin to those in a coronal
hole. The match we obtain between the X-Ray, EUV and magnetogram data and the
numerical results is highly satisfactory concerning overall geometry,
topology of the magnetic field, and density, temperature and velocity of the
plasma. Various basic features obtained through 2D vertical cuts in our
experiments agree qualitatively with the results of
\citet{yokoyama_shibata_96}, in spite of the differences in parameter values.
Beyond all that, our experiments permit discerning for the first
time the 3D geometry of this type of events.  

A number of refinements must be implemented for a better test of the match of
the present numerical results with the observations, like heat conduction or
radiative cooling.  Those improvements are being included into this research
at present.

\acknowledgments
 Financial support by the European Commission through the SOLAIRE
Network (MTRN-CT-2006-035484) and by the Spanish Ministery of Education
through project AYA2007-66502 is gratefully acknowledged, as are the
computer resources, technical expertise and assistance provided by the
MareNostrum (BSC/CNS, Spain) and LaPalma (IAC/RES, Spain)
supercomputer installations. Support from the NASA/{\it Hinode} program and from
the Velux foundation is also acknowledged.  {\it Hinode} is a Japanese mission
developed and launched by ISAS/JAXA, with NAOJ, NASA and STFC as partners, and
co-operated with ESA and NSC. {\it SOHO} is a project of cooperation between ESA
and NASA.

\clearpage

\end{document}